\documentclass[aps,pra,twocolumn,letterpaper,superscriptaddress,10pt]{revtex4-1}
\usepackage{amssymb,amsthm,amsmath,amsfonts}
\usepackage{graphicx,ulem,mathptmx,textcomp}
\usepackage[pdftex,dvipsnames,usenames]{xcolor}
\usepackage[colorlinks=true,urlcolor=blue,citecolor=blue,linkcolor=blue]{hyperref}

\begin{document}

%%%%%%%%%%%%%%%%%%%%%%%%%%%%%%%%%%%%%%%%%%%%%%%%%%%%%%%%
% General Information
%%%%%%%%%%%%%%%%%%%%%%%%%%%%%%%%%%%%%%%%%%%%%%%%%%%%%%%%
\title{Climbing the Fock ladder: Advancing multiphoton state generation}

\author{M. Engelkemeier}
\affiliation{Integrated Quantum Optics Group, Institute for Photonic Quantum Systems (PhoQS), Paderborn University, Warburger Straße 100, 33098 Paderborn, Germany}

\author{J. Sperling}
\email{jan.sperling@upb.de}
\affiliation{Integrated Quantum Optics Group, Institute for Photonic Quantum Systems (PhoQS), Paderborn University, Warburger Straße 100, 33098 Paderborn, Germany}

\author{J. Tiedau}
\affiliation{Integrated Quantum Optics Group, Institute for Photonic Quantum Systems (PhoQS), Paderborn University, Warburger Straße 100, 33098 Paderborn, Germany}

\author{S. Barkhofen}
\affiliation{Integrated Quantum Optics Group, Institute for Photonic Quantum Systems (PhoQS), Paderborn University, Warburger Straße 100, 33098 Paderborn, Germany}

\author{I. Dhand}
\affiliation{Institut f\"ur Theoretische Physik and Center for Integrated Quantum Science and Technology (IQST), University of Ulm, 89069 Ulm, Germany}

\author{M. B. Plenio}
\affiliation{Institut f\"ur Theoretische Physik and Center for Integrated Quantum Science and Technology (IQST), University of Ulm, 89069 Ulm, Germany}

\author{B. Brecht}
\affiliation{Integrated Quantum Optics Group, Institute for Photonic Quantum Systems (PhoQS), Paderborn University, Warburger Straße 100, 33098 Paderborn, Germany}

\author{C. Silberhorn}
\affiliation{Integrated Quantum Optics Group, Institute for Photonic Quantum Systems (PhoQS), Paderborn University, Warburger Straße 100, 33098 Paderborn, Germany}

\date{\today}

\begin{abstract}
	A scheme for the enhanced generation of higher photon-number states is realized, using an optical time-multiplexing setting that exploits a parametric down-conversion source for an iterative state generation.
	We use a quantum feedback mechanism for already generated photons to induce self-seeding of the consecutive nonlinear process, enabling us to coherently add photons to the light that propagates in the feedback loop.
	The addition can be carried out for any chosen number of round trips, resulting in a successive buildup of multiphoton states.
	Our system is only limited by loop losses.
	The looped design is rendered possible by a carefully engineered waveguide source that is compatible with and preserves the shape of the propagating mode.
	We compare the fidelities and success probabilities of our protocol with the common direct heralding of photon-number states.
	This comparison reveals that, for same the fidelity, our feedback-based setup significantly enhances success probabilities, being vital for an efficient utilization in quantum technologies.
	Moreover, quantum characteristics of the produced states are analyzed, and the flexibility of producing higher photon-number states with our setup beyond the common direct heralding is demonstrated.
\end{abstract}

\maketitle

%%%%%%%%%%%%%%%%%%%%%%%%%%%%%%%%%%%%%%%%%%%%%%%%%%%%%%%%
% Introduction
%%%%%%%%%%%%%%%%%%%%%%%%%%%%%%%%%%%%%%%%%%%%%%%%%%%%%%%%
\paragraph*{Introduction.---}

	The photon plays a central role in quantum optics as it represents the fundamental excitation of a quantized light field.
	It is also the basic carrier of quantum information in many quantum communication protocols \cite{BB84,KMNRDM07,NC00}.
	Thus, sources of single photons enable insight into fundamental physics of elementary particles and satisfy a practical demand in quantum technologies.
	These features naturally extend to higher photon numbers.
	Multiphoton states are also prerequisites for even more complex states, such as Holland-Burnett states \cite{HB93}, cat states \cite{OJTG07,MPKPB19}, tensor network states \cite{ZM05,ZPM02,DESBSP18}, interesting multiphoton-entangled states \cite{SPBS19}, which support applications in quantum information science \cite{GT07,KLM01,KMNRDM07,VKLNFGMDS13,BFV09}.
	Therefore, a plethora of experiments investigate what the most efficient generation methods for high-quality single- and multiphoton states are; see, e.g., Refs. \cite{DMPS00,S05,BDSW03,MK09,GFM14, BBFKT19,ZVB04,YMKLF13,NALDT15,CWSS13}. 

	A key element for photon-number (likewise, Fock) state generations are parametric down-conversion (PDC) processes, being a widely accessible and tunable tool.
	Nowadays, a PDC-based heralded generation can be reliably achieved where fidelities above $90\%$ for single- and multiphoton states have been reported \cite{CWSS13,BBFKT19, WDY06}.
	Note that other single-photon sources typically lack the potential to produce higher photon numbers.
	Yet the higher the photon number $n$, the more interesting the applications become.
	However, this is at the expense of an ever increasing demand of resources, e.g., numbers of multiplexed sources and increased measurement times.

	The main limitation in existing PDC-based schemes for generating photon-number states is that the success probability is rather low, significantly diminishing the versatility of such sources \cite{TBHLNGS19,MSM20}.
	This is caused by fundamental limits of generation probabilities, $p$, implying an exponential decay, $p^n$, for generating $n$ photons.
	A simple way to increase $p$ is to increase the intensity of the pump pulse of the PDC process.
	However, unwanted noise contributions in the form of higher (larger than $n$) photon-number components decrease the fidelity with the targeted $n$-photon state considerably.
	Hence, an unavoidable trade-off has to be made between measurement time, set by the upper bounded generation probability, and the state fidelity.
	It constitutes a timely challenge to overcome this limitation, which we do in this contribution.

	Alternative approaches to increase the success probability for producing higher photon-number states are urgently required, while maintaining practicability and compatibility with existing devices.
	Different proposals seek enhancing the generation by making use of a large-scale time multiplexing \cite{KK19}, a coherent-state seed input in the PDC process \cite{S05}, a recycling of the PDC sources several times \cite{MK09}, employing a cavity-based PDC processes \cite{GFM14}, and interference between coherent and Fock states via quantum catalysis \cite{BDSJBDSW12}.
	Inspired by such attempts, we recently put forward a theoretical proposal in which a PDC source in a time-multiplexing architecture with looped configuration is used, together with a heralding that employs multiplexed single-photon detectors \cite{ELSDBDPS20}. 

\begin{figure*}
	\includegraphics[width=\textwidth]{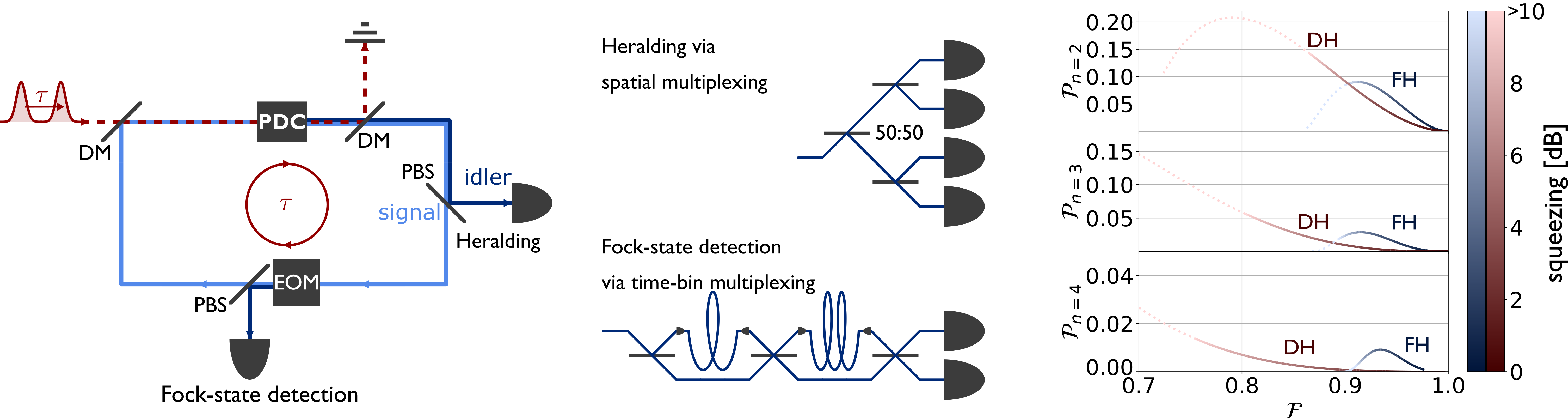}
	\caption{%
		Left:
		Setup overview.
		A laser source optically pumps a PDC source and dichroic mirrors (DMs) ensure that the pump does not interact with the loop.
		Generated photon pairs are separated on a polarizing beam splitter (PBS).
		The vertically polarized idler photons are used for the state heralding.
		Horizontally polarized signal photons continue to propagate through the loop and overlap with subsequent pump pulses in the PDC, controlled by matching the laser's repetition rate with the round-trip time, both $\tau$.
		Thus, the cycling modes seeds the PDC process and more photons are added.
		After a selected number of round trips, the generated signal can be detected through deterministic out-coupling, using an electro-optic modulator (EOM) and a PBS.
		Center:
		Spatially multiplexed detection (top) for the herald with four single-photon counters and time-bin multiplexing (bottom) for the signal state with eight detection time bins.
		In both cases, the incident light is uniformly split into multiple outputs and the joint number of clicks is recorded, approximating a photon-number resolution.
		Right:
		Theoretical prediction of the quantitative relation between fidelity $\mathcal F$ and success probability $\mathcal P$ for the lossless case.
		We consider the generation for $n=2$ (top), $n=3$ (middle), and $n=4$ (bottom) via the common direct heralding (DH, red) from a PDC source and our feedback heralding (FH, blue) approach by varying the PDC pump strength \cite{comment:Squeezing} between a high power (lower $\mathcal F$ values and high squeezing) and zero ($\mathcal F\to1$ and no squeezing).
	}\label{fig:Setup}
\end{figure*}

	In this contribution, we realize the theoretically proposed source and report on the experimental characterization of the states that are thereby generated. 
	We demonstrate a successive buildup of multiphoton states by introducing a time-multiplexing architecture.
	For our conceptual demonstrations, multiphoton states are produced, and a comparison with a standard PDC source is carried out.
	We show a clear enhancement in the success probability of our unconventional approach over stat-of-the-art heralding techniques for multiphoton states with identical fidelities.
	Moreover, this enhancement becomes more pronounced the higher the generated photon number is.
	We additionally analyze the nonclassicality of the generated state and show that we have access to exponentially higher photon numbers than without feedback.
	In conclusion, an efficient iterative generation of multiphoton states is implemented with a platform that is readily available for applications in quantum technologies. 

%%%%%%%%%%%%%%%%%%%%%%%%%%%%%%%%%%%%%%%%%%%%%%%%%%%%%%%%
% Concept
%%%%%%%%%%%%%%%%%%%%%%%%%%%%%%%%%%%%%%%%%%%%%%%%%%%%%%%%
\paragraph*{Concept.---}

	Our experiment is outlined in Fig. \ref{fig:Setup}, left.
	In essence, we make use of a seeded PDC process with heralding to approximate an iterative photon-addition protocol, resulting in increments $\hat{a}^\dag |n\rangle\propto |n+1\rangle$.
	A theoretical analysis of our scheme, including imperfections, can be found in Ref. \cite{ELSDBDPS20}.

	A main innovation compared to standard schemes for multiphoton states is the loop architecture around a PDC source, comparable to a laser cavity.
	The second-order nonlinearity is a type-II PDC process that leads to polarization-non-degenerate photon pairs.
	Vertically polarized photons are immediately directed to a detection unit with a resolution of up to four photons, certifying that this number of photons has been added to the horizontal mode.
	This first step in the iteration scheme is identical to the common approach to producing multiphoton states, named direct heralding (DH) hereafter.

	In our setup, however, horizontally polarized photons (i.e., the signal) can further propagate through the loop and temporally overlap with the subsequent pump pulse, therefore inducing a seeding via quantum feedback of the nonlinear process.
	In this manner, additional signal photons can be coherently added to the cycling mode, similar to the first step with an addition to vacuum.
	A key difference to a standard laser cavity is that our feedback-based system is controllable in terms of number of round trips and pump pulses, ensured by a deterministic out-coupling via an electro-optic modulator.

	The herald and the signal detection consist of a spatial and time-bin multiplexing scheme, respectively \cite{PTKJ96,ASSBWFJPF04}; see Fig. \ref{fig:Setup}, center.
	In both detection schemes, light is successively split on $50{:}50$ beam splitters and each output is measured with a single-photon detector, whose response to detected light is colloquially referred to as a click.
	The number of joint clicks then approximates the number of detected photons \cite{SVA12}.
	Measuring $n$ clicks in the herald detection, having a resolution of up to four clicks, over a given number of round trips then indicates the heralded generation of a photon state $|n\rangle$.
	The signal is characterized using click-counting measurements, however, with a higher resolution of up to eight clicks. 

	The right plot in Fig. \ref{fig:Setup} shows theoretical predictions for the relation of the resulting fidelity $\mathcal F$ to the target $n$-photon state and the probability $\mathcal P$ for successfully generating this state in the lossless scenario.
	(See also Ref. \cite{MK09} in this context.)
	For reasonable squeezing (i.e., pump powers), we predict a significantly enhanced performance of our feedback heralding (FH) when compared with the DH, shown as blue and red curves, respectively.
	Interestingly, the advantage is more pronounced for higher photon numbers (compare plots for $n=2,3,4$ from top to bottom).
	The goal for our implementation is to demonstrate this very enhancement.

%%%%%%%%%%%%%%%%%%%%%%%%%%%%%%%%%%%%%%%%%%%%%%%%%%%%%%%%
% Implementation
%%%%%%%%%%%%%%%%%%%%%%%%%%%%%%%%%%%%%%%%%%%%%%%%%%%%%%%%
\paragraph*{Implementation.---}

	We use type-II PDC in a periodically poled potassium titanyl phosphate waveguide, converting $775~\mathrm{nm}$ light to around $1550~\mathrm{nm}$ telecom wavelength.
	This source is custom-designed and carefully engineered, being spatially single-mode and decorrelated \cite{ECMS11}.
	Importantly, the source exhibits a high spatial mode overlap between in- and out-coupled modes of the waveguide of $96\%$, rendering iterative passes through the source in our scheme possible.
	Furthermore, our source shows a very high single-photon purity, measured by heralded correlations with ${g_{h}}^{(2)}(0) = 0.001$.
	The transmission losses are $0.499~\mathrm{dB/cm}$ and $0.399~\mathrm{dB/cm}$ for horizontally and vertically polarized light, respectively.
	The conversion efficiency for the second-harmonic generation for our source is $1.10\%$.
	
	The source is pumped by a Ti:sapphire laser with a repetition rate of $76~\mathrm{MHz}$.
	The repetition rate corresponds to a pulse separation of $\tau=13.15~\mathrm{ns}$ and is identical to the round trip time $\tau$ in the loop.
	To probe the dependence on the pump power, we used different pulse energies for our measurements that are fitted to squeezing parameters $|\zeta|$ from $\sim0.1$ to $\sim0.3$, likewise $1.0~\mathrm{dB}$--$2.6~\mathrm{dB}$ squeezing \cite{comment:Squeezing}.

	For the detection, we use superconducting nanowire single-photon detectors:
	$N'=4$ detectors with dead times of $10~\mathrm{ns}$ and efficiencies of $75\%$ for the idler and two detectors (for the four time bins in two outputs, resulting in $N=2\times4=8$ detection bins) with $60~\mathrm{ns}$ dead time and $95\%$ efficiency for the signal.
	The overall detection efficiency of the spatial multiplexing is estimated as $\eta'=0.36$, and it is $\eta=0.38$ for the time-bin multiplexing.
	The round-trip losses without the source are $3\%$.
	When including the source, we obtain a loop efficiency $\eta_\mathrm{loop}$ between $50\%$ and $60\%$, being the main limitation.
	This total loop efficiency contains the transmission losses in the waveguide, impurities of the mode overlap, and losses due to all other components in the setup.

	The source is optically pump $t$ times, allowing the signal to propagate $t-1$ times in the loop.
	We monitor the number of heralding clicks, resulting in a click pattern $(k_1,\ldots,k_t)$, where $k_j\in\{0,\ldots,N'\}$ is the number of clicks in the $j$th round.
	Then, the light is coupled out and the click-counting distribution $c_k$ of the signal is measured.
	These data are stored and processed as follows.
	A conditioning to a given pattern, $(k_1,\ldots,k_t)$, determines the targeted number of photons, $n=k_1+\cdots+k_t$.
	Specifically, $t=1$ and $k_1=n$ realizes a DH, and $t=n$ and $(k_1,\ldots,k_t)=(1,\ldots,1)$ implements a FH of an $n$-photon state.
	The success probability $\mathcal P$ is defined as the ratio of measurements that lead to a desired click pattern to the total number of measurements.
	Fidelities are obtained through the Bhattacharyya coefficient, $\mathcal F=\sum_{k=0}^{N}\sqrt{c_k^\mathrm{(sim)}}\sqrt{c_k}$, between the measured data $c_k$ and the simulated click-counting distribution $c_k^\mathrm{(sim)}$ for the target state.
	Moreover, the nonclassicality in terms of the negativity $\mathcal N$ of the matrix of moments of the measured click-counting statistics is determined \cite{SVA13,SBVHBAS15}.

	See the Supplemental Material (SM) \cite{comment:SupplementalMaterial} for further technical details. 
	Also note that our entire data analysis framework is based on measured quantities alone.

%%%%%%%%%%%%%%%%%%%%%%%%%%%%%%%%%%%%%%%%%%%%%%%%%%%%%%%%
% Results
%%%%%%%%%%%%%%%%%%%%%%%%%%%%%%%%%%%%%%%%%%%%%%%%%%%%%%%%
\paragraph*{Results.---}

\begin{figure}[t]
	\includegraphics[width=\columnwidth]{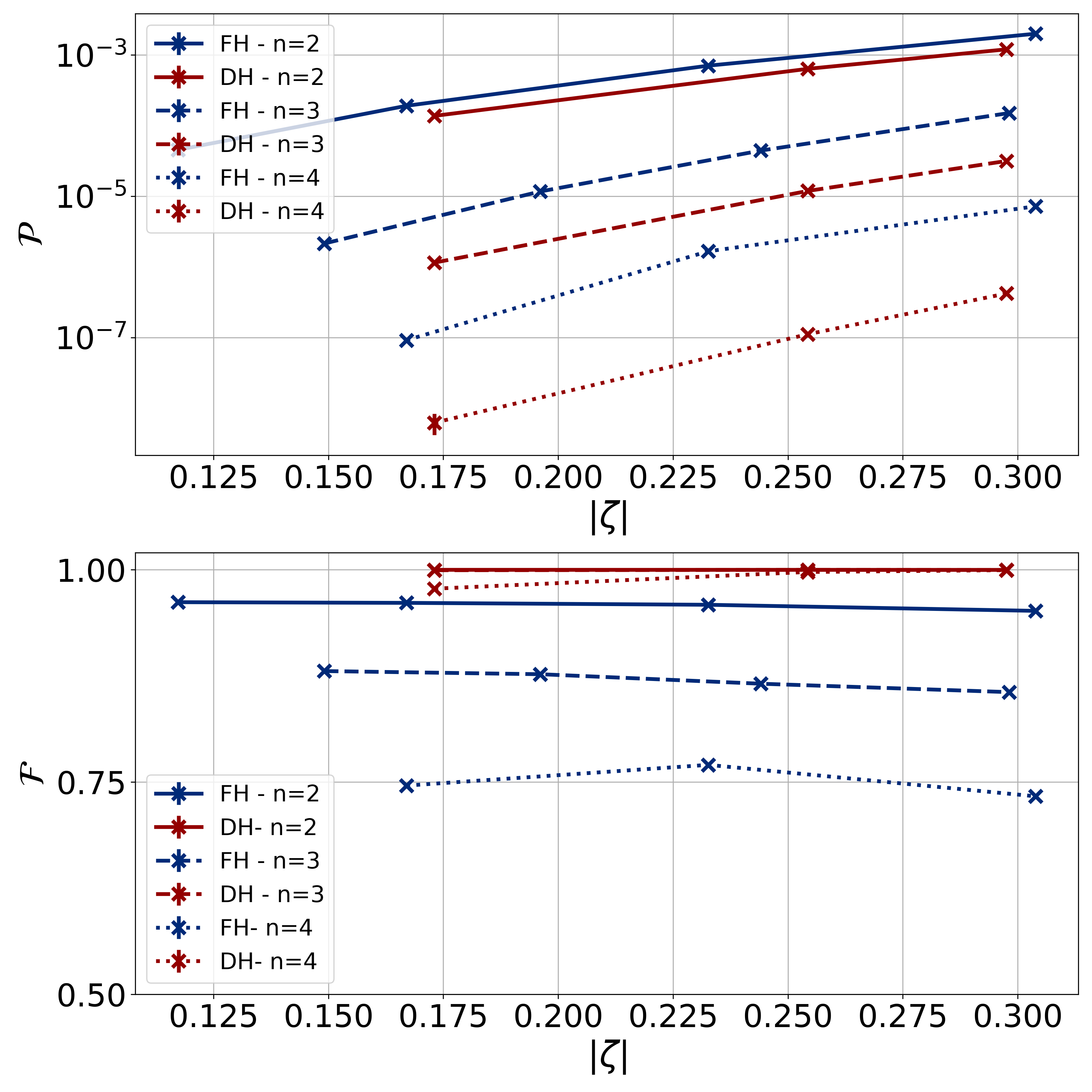}
	\caption{%
		Success probability $\mathcal{P}$ (top) and fidelity $\mathcal{F}$ (bottom)---the former on a logarithmic scale---as a function of the squeezing parameter $|\zeta|$ for $n=2,3,\text{ and }4$ photons (solid, dashed, and dotted, respectively) generated via DH (red) and FH (blue).
		Note that $\mathcal F$ for DH and $n=2\text{ and }3$ (solid and dashed) are almost identical.
	}\label{fig:FPzeta}
\end{figure}

	Figure \ref{fig:FPzeta} shows the measured (crosses) success probabilities (top) and the corresponding fidelities (bottom) for DH (red) and FH (blue) by varying the pulse energy, i.e., the squeezing parameter.
	Uncertainties are included as vertical bars but mostly not visible.
	Lines connect the results for $n=2,3,\text{ and }4$ heralded photons (solid, dashed, and dotted, respectively).
	Note that $c_k^\mathrm{(sim)}$ for obtaining the fidelity $\mathcal F$ here includes the signal detection losses $\eta$ and herald detection losses $\eta'$ as they are identical for both DH and FH.

	The success probability (Fig. \ref{fig:FPzeta}, top) increases with higher pulse energies and is generally higher for smaller $n$ values.
	Importantly, $\mathcal P$ for FH always stays above the corresponding values for the DH.
	In all cases, the success probability increases with the pump power since photon pairs are produced---and thus heralded---at a higher rate.
	For FH and DH, a lower $n$ leads to a higher $\mathcal P$ in an almost equidistant manner in the logarithmic depiction, an expected result as the generation efficiency roughly scales exponentially with the number of photons.
	For a two-photon state, the success probability for FH is as much as $60\%$ higher than the one for the DH scheme.
	By adding further passes through the pumped PDC source, this improvement over the DH continues to increase, resulting in 17-fold increase for the four-photon state.

\begin{figure}[b]
	\includegraphics[width=\columnwidth]{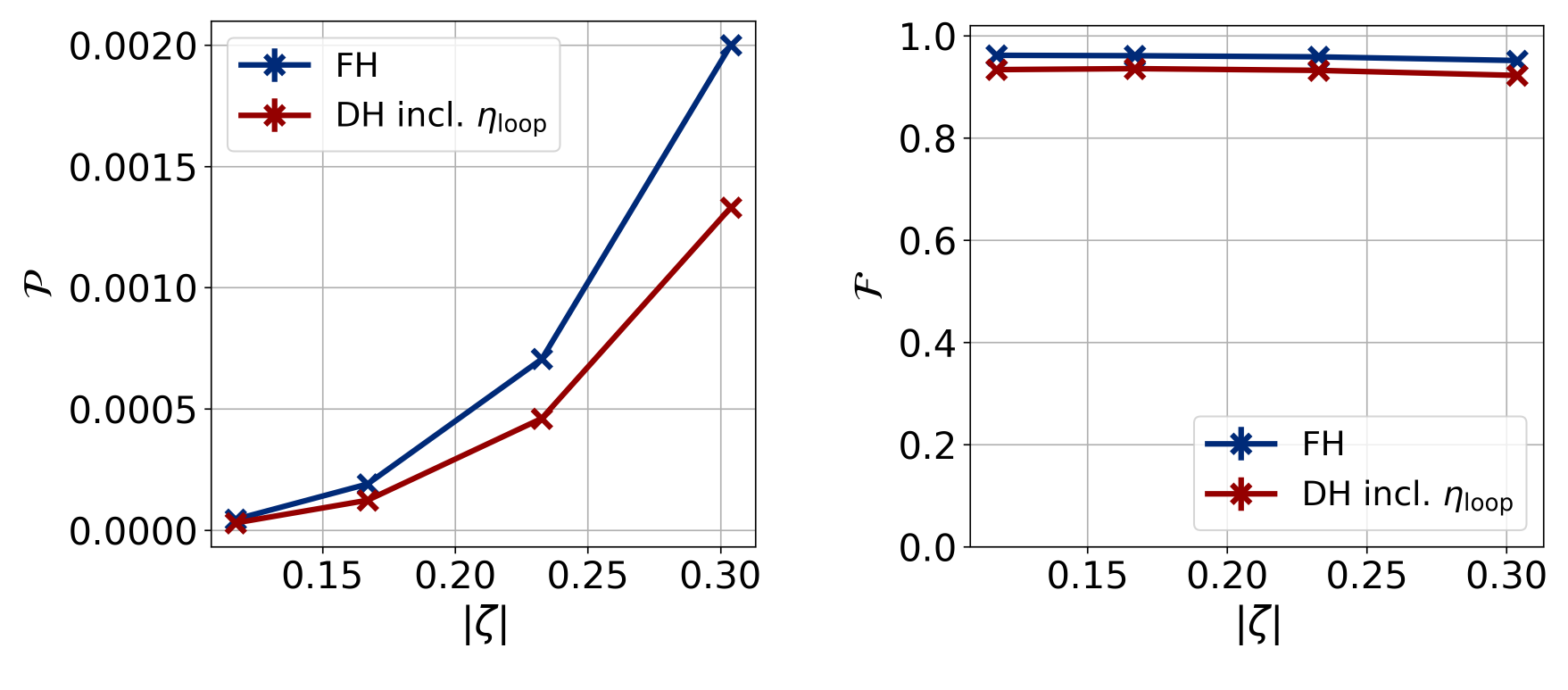}
	\caption{%
		Success probability $\mathcal{P}$ (left) and fidelity $\mathcal{F}$ (right) for the generation of a two-photon state with FH (blue) and DH (red) for the same total losses.
		The latter is realized by propagating a two-photon state from DH in the loop without adding further photons.
		The fidelity is calculated between the measured data and the simulations of ideal two-photons states, including detection losses.
	}\label{fig:FP_extra}
\end{figure}

	The fidelities (Fig. \ref{fig:FPzeta}, bottom) for the DH stay close to $99\%$ while the fidelity for the FH drops with the number of round trips;
	thus, the loop efficiency requires future improvements.
	To account for this fact with the currently available setup, we also compared the DH signal when it is subjected to identical loop losses; see Fig. \ref{fig:FP_extra} for the two-photon generation and the SM \cite{comment:SupplementalMaterial} for the three-photon scenario.
	This is achieved by leaving the DH signal in the loop for an additional round trip, resulting in the heralding click pattern $(2,0)$; recall that the corresponding FH is determined by the pattern $(1,1)$.
	Then, the FH exceeds the DH in both figures of merit, $\mathcal P$ and $\mathcal F$, proving that the current limitation is the loop efficiency $\eta_\mathrm{loop}$.

\begin{figure}[b]
		\includegraphics[width=\columnwidth]{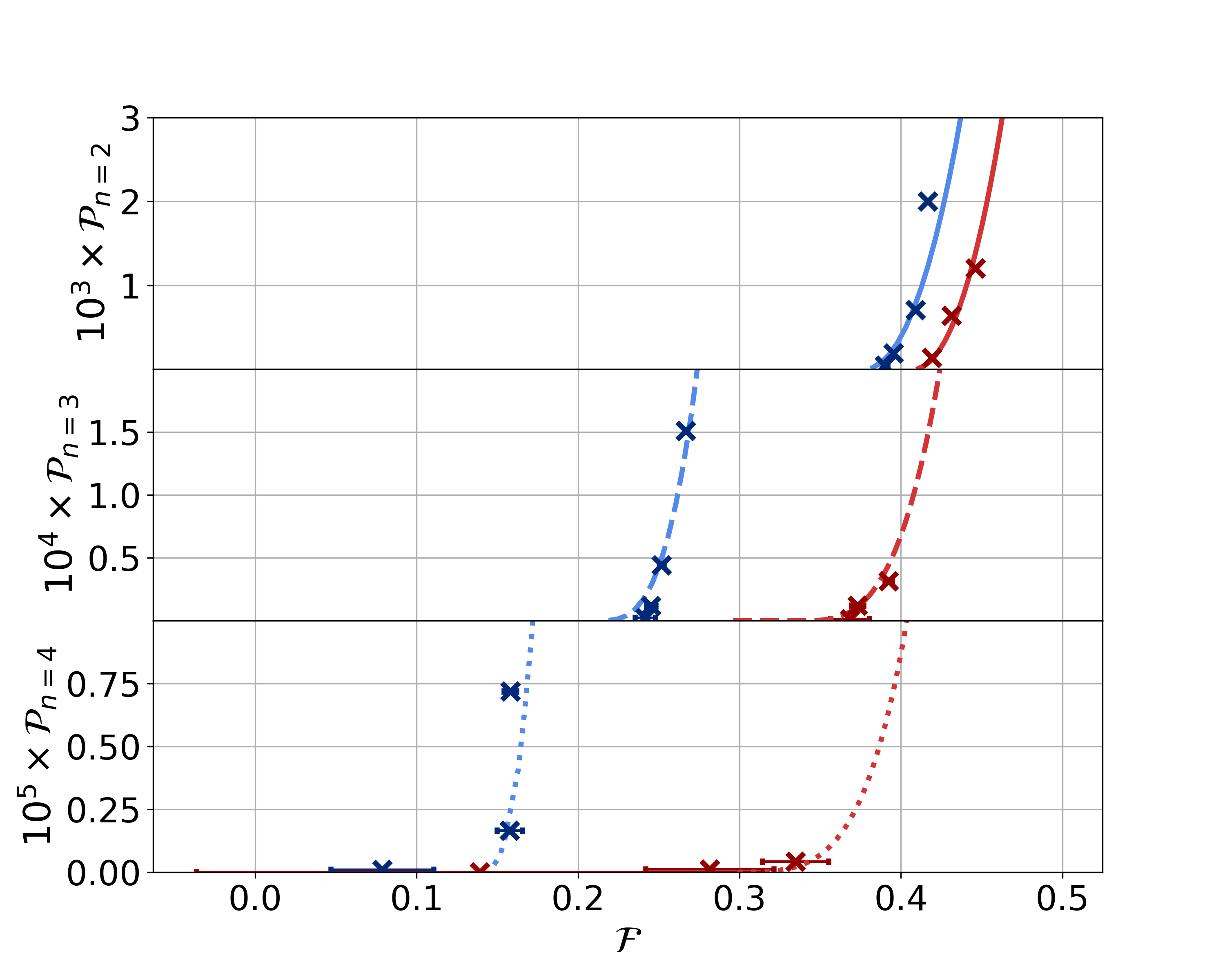}
		\caption{%
			Relation between fidelity $\mathcal{F}$ and success probability $\mathcal{P}$ for DH (dark red) and FH (dark blue) for the generation of $n=2$ (top, solid), $n=3$ (middle, dashed), and $n=4$ (bottom, dotted) photons.
			The light red and blue curves show the theoretical predictions that model the system's imperfections \cite{ELSDBDPS20,comment:SupplementalMaterial}.
		}\label{fig:FP_stirling}
\end{figure}

	A $\mathcal F$-$\mathcal P$ diagram is plotted in Fig. \ref{fig:FP_stirling}, analogously to Fig. \ref{fig:Setup}.
	Note that the changes in the monotonic behavior in the accessible regions of both plots are due to losses.
	The fidelity is reduced here because the comparison of the raw data $c_k$ with a lossless theory $c_k^\mathrm{(sim)}$, i.e., perfect states $|n\rangle$ and detectors.
	The FH clearly outperforms the DH since the former curve is above the latter;
	and only FH can achieve high generation probabilities.
	Hence, the loss of fidelity is less severe than what one gains in success probability.

	Finally, we benchmark the amount of nonclassicality.
	In addition, we demonstrate the versatility of our approach that employs a multiplexed detection for the heralding.
	See Table \ref{tab:NC} for an overview.
	Since we have eight detection bins for the signal, our nonclassicality characterization enables us to employ moments up to the eighth order \cite{SVA12,SBVHBAS15}, exceeding commonly applied second-order correlation functions.

\begin{table}[t]
	\caption{%
		Nonclassicality $\mathcal{N}$, verified through a negative value, and its statistical significance (rightmost column) for different heralding click patterns (first column), resulting in different $n$-photon states, obtained from $t$ pump pulses for a squeezing parameter $|\zeta|\approx0.3$. 
	}\label{tab:NC}
	\begin{tabular}{p{.3\columnwidth}p{.05\columnwidth}p{.05\columnwidth}p{.3\columnwidth}p{.2\columnwidth}}
		\hline\hline
		Heralding pattern & $n$ & $t$ & $\mathcal{N} \times 10^{-4}$  & $|\mathcal{N}|/\sigma(\mathcal{N})$\\
		\hline
		$(2)$ & 2 & 1 & $-13.515$\phantom{0}$\,\pm\,$\phantom{0}$0.096$ & $141$
		\\
		$(1,1)$ & 2 & 2 & \phantom{0}$-9.054$\phantom{0}$\,\pm\,$\phantom{0}$0.050$ & $183$
		\\
		$(2,0)$ & 2 & 2 & \phantom{0}$-4.136$\phantom{0}$\,\pm\,$\phantom{0}$0.048$ & \phantom{0}$87$
		\\
		$(3)$ & 3 & 1 & $-18.3$\phantom{000}$\,\pm\,$\phantom{0}$1.1$ & \phantom{0}$17$
		\\
		$(1,1,1)$ & 3 & 3 & \phantom{0}$-5.35$\phantom{00}$\,\pm\,$\phantom{0}$0.23$ & \phantom{0}$24$
		\\
		$(4)$ & 4 & 1 & $-25$\phantom{.0000}$\,\pm\,$$13$ & \phantom{00}$2$
		\\
		$(2,2)$ & 4 & 2 & $-13.2$\phantom{000}$\,\pm\,$\phantom{0}$2.6$ & \phantom{00}$5$
		\\
		$(1,1,1,1)$ & 4 & 4 & \phantom{0}$-5.2$\phantom{000}$\,\pm\,$\phantom{0}$1.1$ & \phantom{00}$5$
		\\
		$(1,1,1,2)$ & 5 & 4 & $-16.9$\phantom{000}$\,\pm\,$\phantom{0}$7.0$ & \phantom{00}$2.4$
		\\
		$(1,1,1,3)$ & 6 & 4 & $-63$\phantom{.0000}$\,\pm\,$$37$ & \phantom{00}$1.7$
		\\
		\hline\hline
	\end{tabular}
\end{table}

	The first three rows of the table compare the nonclassicality for two-photon states, in the order: DH, FH, and DH with loop losses.
	While the mean nonclassicality $\mathcal N$ for DH is the highest, it is generated with a statistical significance (i.e., reliability) which is lower than for FH.
	When taking loop losses into account, as discussed above, the FH beats the DH by more than a factor of two in the absolute negativity, too.
	For the three-photon generation, fourth and fifth row, the significance is again higher for the FH (24 standard deviations) compared with the DH (17 standard deviations).
	For the four-photon generation, rows with $n=4$, we additionally include the case of generating two photons in each of two round trips.
	Interestingly, this results in a four-photon state with a significance that is identical to the FH case and an absolute negativity of the same order of magnitude as in the DH scenario. 

	As a proof of concept, the remaining rows in Table \ref{tab:NC} exemplify generalized heralding schemes with $n>4$ photons, being impossible with DH with four detectors.
	In principle, we can produce exponentially more photons via our FH, $n\leq (N')^t$, with $N'$ multiplexed heralding detectors and $t$ passes through the source, while DH is constrained to $t=1$.
	Consequently, our FH supersedes the DH in terms of generating the targeted states and allows for a more flexible approach to higher-order photon-number states.
	Unlike the DH, the FH can, in principle, generate arbitrary photon-number states with a finite detector resolution, $N'$, and $t$ is only loop-loss-limited.
	
%%%%%%%%%%%%%%%%%%%%%%%%%%%%%%%%%%%%%%%%%%%%%%%%%%%%%%%%
% Conclusion
%%%%%%%%%%%%%%%%%%%%%%%%%%%%%%%%%%%%%%%%%%%%%%%%%%%%%%%%
\paragraph*{Conclusion.---}

	Going conceptually and performance-wise beyond existing sources, we realized an advanced source for multiphoton states by operating a PDC source in a quantum feedback loop.
	No equivalent source has shown to experimentally generate multiphoton states this manner, rendered possible by an optimized source engineering.
	Rather than using a DH, we operate our PDC source in a time-multiplexing feedback loop, allowing us to coherently add photons with every round trip.
	The coherent addition of more than one photon to the traveling mode is achieved by a spatial multiplexing of four detectors, yielding a quasi-photon-number resolution that is typically not considered in cavity-based heralding schemes.
	A high mode overlap of the deterministically in- and out-coupled light of the PDC source and the high stability of our setup ensures the efficient operation of our system as an iterative photon adder in a loop-based architecture. 

	We showed that our FH exceeds the generation rate of photon-number states when compared to traditional approaches, and this effect is more pronounced for higher photon numbers.
	The success probability for a two-photon state generated with the FH is $60\%$ higher than the one for the DH source, and we found an up to 17-fold improvement for a four-photon state.
	The FH exhibits a high fidelity with perfect photon-number states, limited by the loop losses.
	When accounting for the latter imperfections, an improvement that is worth pursuing in future, we find that our scheme does also result in higher fidelities when compared to the traditional loopless approach.
	The same can be concluded from our nonclassicality analysis in which we also generated states that are inaccessible with a comparable DH. 

	Therefore, an advanced time-multiplexed PDC source for the successive buildup of arbitrary photon-number states has been introduced that overcomes the fundamental limitation of other sources.
	This renders it possible to realize protocols in quantum technology that require such highly nonclassical states of light to overcome limitations of classical systems.

%%%%%%%%%%%%%%%%%%%%%%%%%%%%%%%%%%%%%%%%%%%%%%%%%%%%%%%%
% Akn
%%%%%%%%%%%%%%%%%%%%%%%%%%%%%%%%%%%%%%%%%%%%%%%%%%%%%%%%
\paragraph*{Acknowledgments.---}

	The Integrated Quantum Optics group acknowledges financial support through the European Commission through the ERC project QuPoPCoRN (Grant No. 725366) and the H2020-FETFLAG-2018-03 project PhoG (Grant No. 820365) and funding through the Gottfried Wilhelm Leibniz-Preis (Grant No. SI1115/3-1).
	The Institute of Theoretical Physics group acknowledges financial support by the BMBF through the Q.Link.X and the QR.X projects and the Alexander von Humboldt-Foundation.

%%%%%%%%%%%%%%%%%%%%%%%%%%%%%%%%%%%%%%%%%%%%%%%%%%%%%%%%
% SUPPLEMENTAL MATERIAL
%%%%%%%%%%%%%%%%%%%%%%%%%%%%%%%%%%%%%%%%%%%%%%%%%%%%%%%%
\onecolumngrid
\section*{Supplemental Material}
\twocolumngrid
\appendix

	Here, we provide further theoretical and experimental details in addition to the key results as presented in the main text.
	In Appendix \ref{SM:TB}, we briefly reintroduce the theoretical framework as used for the analysis and simulation.
	In Appendix \ref{SM:error}, the estimation of uncertainties for fidelity, success probability, and nonclassicality is formulated.
	Appendix \ref{SM:ExpDescription} contains a detailed setup description.
	In Appendix \ref{SM:FPanalysis}, further experimental results are presented.
	In Appendix \ref{SM:NCanalysis}, we show additional analyses of the states' nonclassicality.

%%%%%%%%%%%%%%%%%%%%%%%%%%%%%%%%%%%%%%%%%%%%%%%%%%%%%%%%
\section{Lossless scenario}\label{SM:TB}

	For the readers' convenience, we briefly recall the model developed in Ref. \cite{ELSDBDPS20}.
	It is based on exponential operators of the form $\hat E(x)=x^{\hat n}={:}e^{(x-1)\hat n}{:}$ and a two-mode squeezer with a gain $\gamma=\cosh^2 |\zeta|$, where $|\zeta|$ is the amplitude of squeezing parameter, being proportional to the square root of the pump power of a single pulse.
	Then, if the inputs to the two-mode squeezer are vacuum and $\hat E(x)$ and one of the outputs is traced out with $\hat E(z)$, the remaining output reads
	\begin{equation}
		\hat F(x,z)=\frac{1}{\gamma}\hat E\left(\frac{x+[\gamma-1]z}{\gamma}\right).
	\end{equation}

	The positive operator-valued measure of the click-counting statistics takes the form $\hat\Pi_k=\binom{N}{k}\sum_{j=0}^k\binom{k}{j}(-1)^{k-j}\hat E(j/N)$ for $k$ clicks from $N$ detectors, assuming no detection losses and being a linear combination of the previously defined exponential operators.
	For example, for a perfect $n$-photon state, the resulting click-counting distribution simplifies to
	\begin{equation}
		c^{(n)}_k=\langle n|\hat\Pi_k|n\rangle=\binom{N}{k}\frac{k!}{N^n}\begin{Bmatrix}n\\k\end{Bmatrix},
	\end{equation}
	where $\left\lbrace\begin{smallmatrix}n\\k\end{smallmatrix}\right\rbrace=k!^{-1}\sum_{j=0}^k\binom{k}{j}(-1)^{k-j}j^n$ denotes the Stirling number of the second kind.
	With this approach from Ref. \cite{ELSDBDPS20}, we can model the DH and FH, including imperfections, such as losses, (excess) noise contributions, and the finite detector resolution.

	In a first application, we consider the ideal case of direct heralding.
	Then, we get as the output of the two-mode squeezer with two vacuum inputs ($x=0$) for a measurement of $n$ clicks from $N'$ heralding detectors the following, not normalized state:
	\begin{equation}
		\hat\rho_n=\frac{1}{\gamma}\binom{N'}{n}\sum_{j=0}^n \binom{n}{j}(-1)^{n-j}\hat E\left(\frac{[\gamma-1]j}{\gamma N'}\right),
	\end{equation}
	being a linear combination of our exponential operators that originates from the expansion of the click-counting measurement operators, specifically $\hat\Pi_n$.
	The trace of this operator then yields the normalization and is identical to the success probability,
	\begin{equation}
		\mathcal P=\mathrm{tr}(\hat\rho_n)=\frac{1}{\gamma}\binom{N'}{n}\sum_{j=0}^n \binom{n}{j}\frac{(-1)^{n-j}\gamma N'}{\gamma N'-[\gamma-1]j}.
	\end{equation}
	Also, the normalized click-counting statistics for $k$ clicks from $N$ multiplexed detectors for this state can be directly obtained,
	\begin{equation}
	\begin{aligned}
		c^\mathrm{(DH)}_{k}=&\frac{1}{\mathcal P\gamma}\binom{N'}{n}\binom{N}{k}\sum_{j=0}^n\sum_{j'=0}^{k} \binom{n}{j}\binom{k}{j'}
		\\
		&\times
		(-1)^{n-j}(-1)^{k-j'}\frac{NN'\gamma}{NN'\gamma-[\gamma-1]jj'}.
	\end{aligned}
	\end{equation}
	This expression can be used to compute the fidelity with the click-counting statistics of the perfect $n$-photon state, $\mathcal F=\sum_{k=0}^N\sqrt{c_k^{(n)}c_k^\mathrm{(DH)}}$.
	The resulting figures for $\mathcal P$ and $\mathcal F$ are plotted in Fig. 1 (right) of the main text to illustrate the theory of DH with multiplexing detectors.

	The second example concerns the iterative addition of one photon over $n$ round trips, initialized with vacuum, i.e., the FH scenario in the same Fig. 1.
	For this purpose, it is convenient to explicitly write $\hat\Pi_1=N'[\hat E(1/N')-\hat E(0)]$, where $\hat E(0)=|0\rangle\langle 0|$, for one click from $N'$ multiplexed on-off detectors.
	Thus, for the single input $\hat E(x)$, we find the output of one heralding click as $[N'/\gamma]\hat E(\gamma^{-1})[\hat E(x+\tfrac{\gamma-1}{N})-\hat E(x)]$.
	The $n$-fold iteration with an initial value $x=0$ then results in the not normalized state
	\begin{equation}
		\hat\rho_n=
		\left(\frac{N'}{\gamma}\right)^n
		\hat E\left(\frac{1}{\gamma^n}\right)
		\sum_{j=0}^n\binom{n}{j}(-1)^{n-j}\hat E\left(j\frac{[\gamma-1]}{N'}\right).
	\end{equation}
	As in the previous example, the normalization yields the success probability and reads
	\begin{equation}
		\mathcal P=\mathrm{tr}(\hat\rho_n)=
		\left(\frac{N'}{\gamma}\right)^n
		\sum_{j=0}^n\binom{n}{j}
		\frac{(-1)^{n-j}N'\gamma^n}{N'\gamma^n-[\gamma-1]j}.
	\end{equation}
	The click-counting statistics takes, from which one can determine the fidelity $\mathcal F=\sum_{k=0}^N\sqrt{c_k^{(n)}c_k^\mathrm{(FH)}}$, obeys
	\begin{equation}
	\begin{aligned}
		c_k^\mathrm{(FH)}=&\frac{1}{\mathcal P}
		\left(\frac{N'}{\gamma}\right)^n\binom{N}{k}
		\sum_{j=0}^n\sum_{j'=0}^k
		\binom{n}{j}\binom{k}{j'}
		\\
		&\times (-1)^{n-j}(-1)^{k-j'}\frac{NN'\gamma^n}{NN'\gamma^n-[\gamma-1]jj'}.
	\end{aligned}
	\end{equation}

	All above considerations already include excess noise from the amplification process that is the two-mode squeezer and the finite resolution of the multiplexing detectors \cite{ELSDBDPS20}.

	For our other simulations, the only thing left is the treatment of loss that can be achieved with a simple modification of the ideal exponential operators, like elaborated on in detail in Ref. \cite{ELSDBDPS20}.
	For instance, signal detection losses $\eta$ are included in the click-counting statistics of the target state $|n\rangle$ for obtaining the simulated $c_k^\mathrm{(sim)}$ that is compared with the measured data via the fidelity in Figs. 2 and 3 in the main text.

%%%%%%%%%%%%%%%%%%%%%%%%%%%%%%%%%%%%%%%%%%%%%%%%%%%%%%%%
\section{Error estimation and propagation}\label{SM:error}

	In the following, we discuss the general error estimation and propagation for the fidelity, success probability, and nonclassicality as applied to obtaining the results in the main text.
	 
	Suppose $C_k$ denotes the click-counting events for the signal using $N=8$ multiplexed detection time bins, where $k$ is the number of detection bins jointly recording a click.
	The total number of recorded events is given by $C=\sum_{k=0}^N C_k$.
	Furthermore, we want to estimate the value of a quantity $f$, described by the parameters $f_k$ for each number $0\leq k\leq N$ of joint clicks, resulting in the mean value
	\begin{equation}
		\overline{f}=\frac{1}{C}\sum_{k=0}^N f_k C_k.
	\end{equation}
	Similarly, the second moment is given by $\overline{f^2}=\sum_{k=0}^N f_k^2 C_k/C$.
	From the above relations, the variance of $f$ can be estimated, which then yields the random error as
	\begin{equation}
		\sigma(f)=\sqrt{\frac{\overline{f^2}-\overline{f}^2}{C-1}}.
	\end{equation}

	Such a direct sampling and error estimation of relevant quantities is an efficient way to obtain experimental values, $f=\overline{f} \pm\sigma(f)$, for linear expressions.
	For instance, the $m$-th normally ordered moment of the click-counting statistics is described through
	\begin{equation}
		f_k=\frac{\binom{k}{m}}{\binom{N}{m}},
	\end{equation}
	where $\binom{k}{m}=0$ for $m>k$ and $0\leq m\leq N$ \cite{SVA13}.
	From these moments, we can determine the matrix of moments $M=\overline{M}\pm \sigma(M)$ to determine nonclassicality \cite{SVA13}, where $\sigma(M)$ is the matrix of uncertainties of the individual moments.
	This is done by finding the normalized eigenvector $v$ to the minimal eigenvalue of $\overline{M}$.
	The mean negativity for nonclassical light is then given by $v^\dag\overline{M}v$ and a quadratic error propagation yields the uncertainty $[(v^2)^\dag \sigma(M)^2(v)^2]^{1/2}$, where ${\bullet}^2$ denotes the entry-wise square;
	see also Ref. \cite{SBVHBAS15} for details.
	The thereby determined negativity is $\mathcal N$, as used in the main text.

	Another example concerns the determination of a random error of the statistics itself, achieved by setting $f_k=\delta_{k,k'}$, where $\delta_{k,k'}=1$ for $k=k'$ and $\delta_{k,k'}=0$ otherwise.
	This yields $\overline{f}=C_{k'}/C=\overline{f^2}$ and $\sigma(f)=[\overline{f}(1-\overline{f})/(C-1)]^{1/2}$.
	Therefore, the estimate of the click-counting statistics reads
	\begin{equation}
		\label{SMeq:estimateCCstatistics}
		c_{k}
		=\overbrace{\overline{c_k}}^{=C_k/C}
		\pm\underbrace{\sqrt{\frac{\overline{c_k}(1-\overline{c_k})}{C-1}}}_{=\sigma(c_k)},
	\end{equation}
	which is useful for the following nonlinear error propagation.
	Also, this expression is helpful when determining an error for success probabilities to measure $k$ heralding clicks in the case that we consider the heralding detection rather than the signal measurements.
	Note that $k=(k_1,\ldots,k_t)$ could be a multi-index in the case of $t$ source passes, with $k_j$ clicks in the $j$th propagation through the source.
	Then the above formula is used to determine errors for the success probability $\mathcal P$ for a given heralding click pattern.

	We now consider a general, nonlinear function $F=F(c_0,\ldots,c_N)$ of the click-counting statistics.
	Employing the standard technique for statistical estimates and error propagation, we get the mean value
	\begin{equation}
		\overline{F}=F(\overline{c_0},\ldots,\overline{c_N})
	\end{equation}
	and, using a quadratic error propagation, the random error estimate
	\begin{equation}
		\sigma(F)=\left[
			\sum_{k=0}^N \left(\frac{\partial F(\overline{c_0},\ldots,\overline{c_N})}{\partial c_k}\right)^2\sigma(c_k)^2
		\right]^{1/2}
	\end{equation}
	to which Eq. \eqref{SMeq:estimateCCstatistics} can be applied for identifying $\overline{c_k}$ and $\sigma(c_k)$.
	For instance, the fidelity is given by a nonlinear expression of the form
	\begin{equation}
		F=\sum_{k'=0}^N F_{k'}\sqrt{c_{k'}},
	\end{equation}
	where $F_k$ is the square root of the target click-counting distribution, obtained theoretically as outlined in Appendix \ref{SM:TB}.
	The derivative then yields $\partial F/\partial c_k=F_k/(2\sqrt{c_k})$.
	Thus, we get $\overline{F}=\sum_{k'=0}^N F_{k'}\overline{c_{k'}}^{1/2}$ and
	\begin{equation}
 		\sigma(F)=\frac{1}{2\sqrt{C-1}}\sqrt{\sum_{k=0}^N F_k^2(1-\overline{c_k})}
	\end{equation}
	when applying the expressions derived previously.
	For $F_k=(c_k^\mathrm{(sim)})^{1/2}$, we obtain the error margins for the fidelity $\mathcal F$.

%%%%%%%%%%%%%%%%%%%%%%%%%%%%%%%%%%%%%%%%%%%%%%%%%%%%%%%%
\section{Additional experimental details and results}

	In this appendix, we provide further technical details about the experiment and additional results from our vast data analysis.
	We show a detailed sketch of the experimental setup used for the state generation, give details about the measurements and the data acquisition.
	Furthermore, we recorded data for $t\in\{1,2,3,4\}$ passes of the source and for different squeezing levels, as provided in Table \ref{tab:zeta}.
	Moreover, we have performed additional analysis of the data regarding fidelities, success probabilities and nonclassicalities.

\begin{table}[t]
	\caption{%
		Realized pump pulses $t$ and the fitted values depending on the pulse energies for their squeezing parameters $|\zeta|$.
	}\label{tab:zeta}
	\begin{tabular}{p{.22\columnwidth}|p{.22\columnwidth}|p{.22\columnwidth}|p{.22\columnwidth}}
		\hline\hline
		$|\zeta|\text{ for }t=1$ &  $|\zeta|\text{ for }t=2$ & $|\zeta|\text{ for }t=3$ & $|\zeta|\text{ for }t=4$
		\\
		\hline
		& 0.1172	& 0.1491	&	\\
		0.1730	& 0.1670	& 0.1961	& 0.1670\\
		0.2543	& 0.2326	& 0.2440	&0.2326	\\
		0.2975	& 0.3038	& 0.2980	&0.3038	\\
		\hline\hline
	\end{tabular}
\end{table}

\subsection{Detailed description of the experiment}\label{SM:ExpDescription}

\begin{figure}[b]
	\includegraphics[width=\columnwidth]{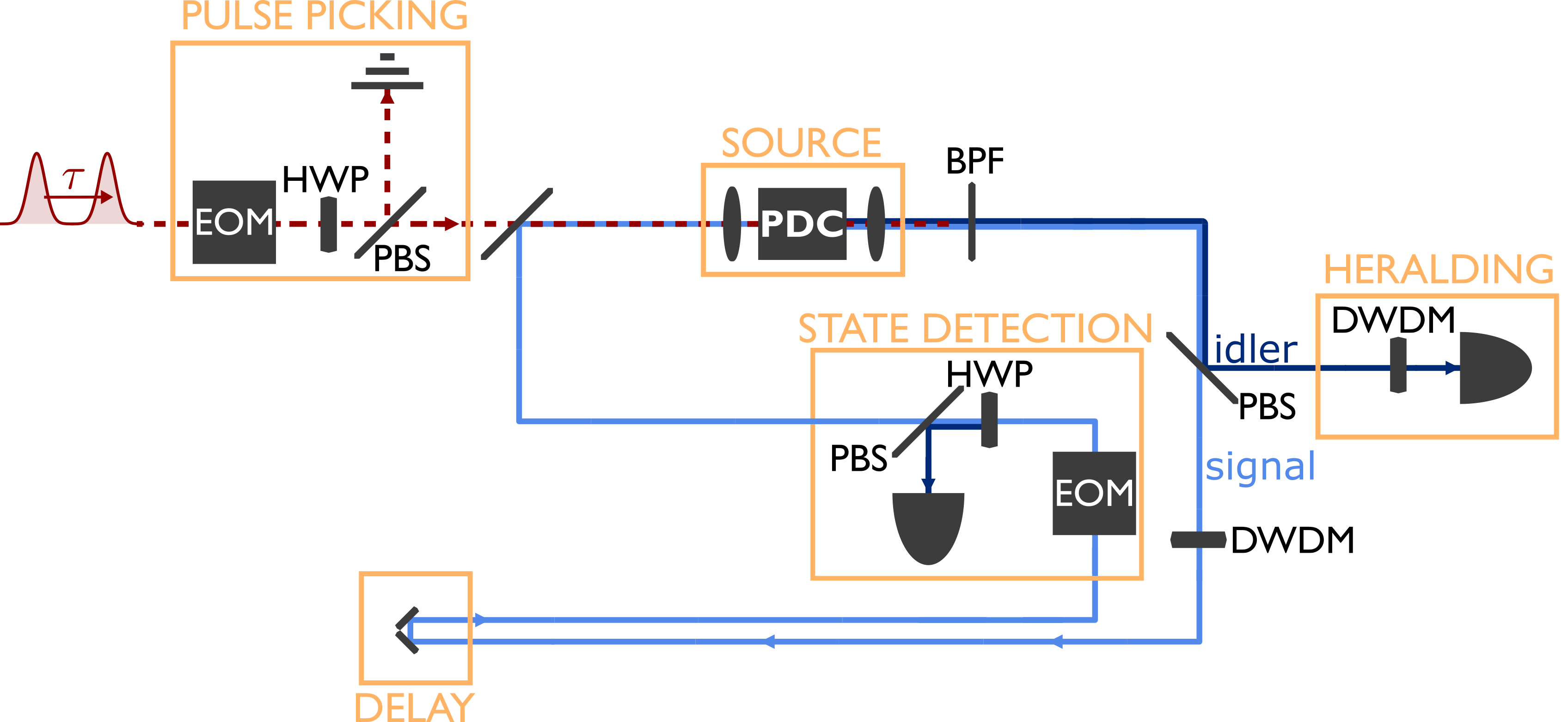}
	\caption{%
		Detailed setup sketch of our approach to generating photon-number states.
		HWP: half-wave plate;
		PBS: polarizing beam spitter;
		PDC: parametric down-conversion;
		BPF: band-pass filter;
		DWDM: dense wavelength division multiplexing;
		EOM: electro-optic modulator.
		For further information, please see the text.
	}\label{fig:detail}
\end{figure}

	Figure \ref{fig:detail} shows a more detailed sketch of our setup.
	There are five parts in the setup that are described in the following:
	pulse picking, source, delay, heralding, and detection.

	First, the measurements for the FH are performed with exactly $n$ pump pulses for generating an $n$-photon state, and we use only one pump pulse entering the setup for the DH.
	Therefore, a pulse picking is realized for sending exactly the wanted number of pulses into the setup.
	This is done by a fast EOM, which changes the polarization of the initial pulses from vertical to horizontal polarization so that only the picked pulses enter the setup at the PBS.
	The HWP is used to match the polarization axes of the EOM and the PBS.

	These pump pulses enter the setup, are coupled in and out of the PDC source via lenses, and are filtered after the source by a BPF.
	The source is a periodically poled potassium titanyl phosphate (ppKTP) chip with rubidium in-diffused waveguides.
	Because of this custom-engineered ppKTP source and waveguides, the photons are spatially single mode, the photon pair is decorrelated, and we have a back-coupling efficiency between the out- and in-coupling of the cycling mode of $96\%$, allowing us to utilize this waveguide for the feedback.

	We use narrow-band DWDM filter in both arms for a spatial filtering of the side peaks in the spectrum of both signal and idler.
	The waveguide has low transmission losses of $0.399~\mathrm{dB/cm}$ for vertical and $0.499~\mathrm{dB/cm}$ for horizontal polarization, which result into a total loss of $8.78\%$ and $10.85\%$, respectively.
	Moreover, the conversion efficiency is $1.10\%$, and the central wavelength within the PDC process is given as $771~\mathrm{nm}$ for the pump and $1542~\mathrm{nm}$ for signal and idler.
	The single-photon purity, given by the heralded quantum correlation function, is ca. $g_h^{(2)}(0)=0.001$ and generally depends on the pulse energies.

	Signal and idler are then separated by propagating thought a PBS, and the idler is detected and functions as the heralding. 
	The heralding is performed by a spatial-multiplexed detection with a resolution of four clicks, where each detector has a efficiency of $76\%$ and a dead time of $10~\mathrm{ns}$.
	
	The signal photon enter the feedback path where we have to match the feedback path length with the repetition rate of $76~\mathrm{MHz}$ of the laser pulses.
	Because of several optical elements in the feedback, such as the crystal in the EOM and the PBS, the effective path length of the feedback is different to the geometric path.
	Hence, we match both in a range less than $1\,\mu\mathrm{m}$, which is accessible by fine-tuning with a delay stage.
	Specifically, the delay defines our overlap of the pump pulses and the feedback mode.

	After a selected number of round trips, the cycling mode is coupled out of the loop, which is performed deterministically by matching the switching of the first and second EOM.
	The polarization of the cycling mode is thus changed into vertical and reflected at the second PBS.
	Again, a HWP is used for matching the polarization axes of the EOM and the PBS.
	The subsequent signal state detection is carried out with a time-multiplexed detection with an eight-click resolution, where the detectors have efficiencies of $95\%$ and dead times of $60~\mathrm{ns}$.

	For the data acquisition, we perform experiments in blocks, in which every block contains $10\,000$ measurements, where a measurement is defined by one initial trigger event every $1~\mathrm{MHz}$.
	Here, we realized $30\,000$ blocks for one data point for every state generation, resulting in a total of 300 million individual measurements.
	With increasing number of time bins in the heralding, depending on the number of pump pulses entering the source, the measurement time is increasing, too.
	The measurement time for one single block for the DH is $0.896~\mathrm{s}$;
	for the two-photon state generation with the FH, the time increases to $1.324~\mathrm{s}$;
	for the three- and four-photon state, it is $1.576~\mathrm{s}$ and $2.060~\mathrm{s}$, respectively.
	Thus, the total measurement for one data point for the DH is $7.49$ hours; for the FH, it is $11.03$ hours for two, $13.13$ hours for three, and $17.16$ hours for four source passes, respectively.

	We used our model from Ref. \cite{ELSDBDPS20} (see also Appendix \ref{SM:TB}) to compare our data with theoretical predictions.
	Our measurements included several pump energies, which we fitted to our model for obtaining several system parameters, such as detection efficiencies (given in the main text) and squeezing parameters (Table \ref{tab:zeta}).
	For better comparison reasons, we carry out the measurements for each state for comparable squeezing parameters.
	As an additional information, the fitted loop-propagation efficiency for $t-1=1$ round trip is $60\%$ and dropped to $55\%$ and $52\%$ for $t=3$ and $t=4$, respectively.

%%%%%%%%%%%%%%%%%%%%%%%%%%%%%%%%%%%%%%%%%%%%%%%%%%%%%%%%
\subsection{Fidelity and success probability}\label{SM:FPanalysis}

\begin{figure}[t]
	\includegraphics[width=\columnwidth]{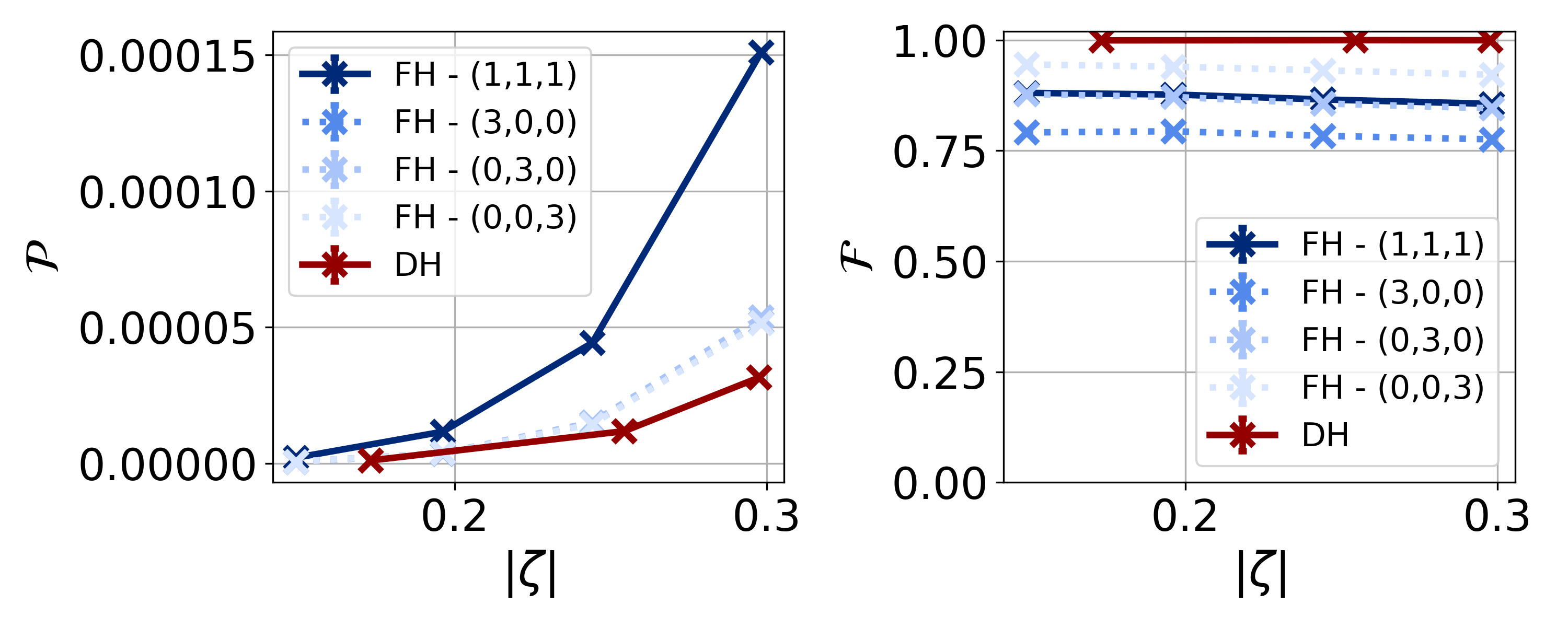}
	\caption{%
		Success probability $\mathcal{P}$ (left) and fidelity $\mathcal F$ (right) for a $n=3$-photon state for DH (red), FH (dark blue), and additional three-photon-heralding click patterns (various shades of blue, dotted).
	}\label{fig:n3comp}
\end{figure}

	Similarly to Fig. 3 for two photons in the main text, Fig. \ref{fig:n3comp} shows the fidelity and the success probability for the three-photon state generation with the DH, the FH, and the additional cases of exactly three heralding clicks in only one round trip, corresponding to the click patterns $(3,0,0)$, $(0,3,0)$, and $(0,0,3)$.
	This is done to compare various DH scenarios of $n=3$ photons that is additionally subject to the same amount of round-trip losses as the FH.

	The fidelity of three clicks in the first and last round trip differ from the state's fidelity with the FH pattern $(1,1,1)$.
	But the pattern with three clicks in the second cycle stays close to the FH.
	The explanation for this behavior is the same as for the special cases in the two-photon state generation;
	different pattern see different amplifications and loop losses.
	Since the case $(0,0,3)$ is not affected by loop loses, the fidelity is the highest while the $(3,0,0)$ case is significantly affected by the loop loss.
	The success probabilities for the additional cases are almost the same, still higher than for the DH.

	For complementing the results shown in form of Fig. 2 in the main text, we provide the corresponding values, including a one-standard-deviation error margin, in Table \ref{tab:T=2} for the two-photon, in Table \ref{tab:T=3} for the three-photon, and in Table \ref{tab:T=4} for the four-photon generation. 
	One can see that the uncertainties for the success probability and the fidelity are increasing with increasing the photon number.
	Furthermore, the uncertainty is increasing faster for the DH compared to the FH. 
	For example, the success probability for the highest squeezing parameter $|\zeta|$ for a four-photon state shows that the uncertainty of the DH case is five times larger than for the FH.
	This strengthens the observation in terms of a statistical quantification that we generate in total more successive events for the given heralding case in the feedback-based system.
	Another point worth mentioning here is that, for both DH and FH, the uncertainty increases when decreasing the squeezing parameter $|\zeta|$ because the number of recorded events is dropping.

\begin{table}[t!]
	\caption{%
		Fidelities and success probabilities, including their uncertainties, for $n=2$ and comparable squeezing parameters $|\zeta|$. 
	}\label{tab:T=2}
	\begin{tabular}{p{.1\columnwidth}p{.15\columnwidth}p{.35\columnwidth}p{.35\columnwidth}}
		\hline\hline
		&	$|\zeta|$  & $\mathcal F~\mathrm{[\%]}$ & $\mathcal P~\mathrm{[\text{\textperthousand}]}$
		\\
		\hline
		FH &0.167 & $96.10 \pm (7.48 \times 10^{-4})$ & $0.191 \pm (2.60 \times 10^{-4})$\\
		DH &0.173 & $99.99 \pm (7.51 \times 10^{-4})$ & $0.137 \pm (3.08 \times 10^{-4})$\\
		\hline
		FH &0.233 & $95.88 \pm (6.09 \times 10^{-4})$ & $0.706 \pm (4.02 \times 10^{-4})$\\
		DH &0.254 & $99.99 \pm (7.73 \times 10^{-4})$ & $0.637 \pm (6.76 \times 10^{-4})$\\
		\hline
		FH &0.304 & $95.17 \pm (7.89 \times 10^{-4})$ & $1.999 \pm (8.59 \times 10^{-4})$\\
		DH &0.298 & $99.98 \pm (7.84 \times 10^{-4})$ & $1.203 \pm (9.41 \times 10^{-4})$\\
		\hline\hline
	\end{tabular}
\end{table}

\begin{table}[t!]
	\caption{%
		Fidelities and success probabilities, including their uncertainties, for $n=3$ and comparable squeezing parameters $|\zeta|$. 
	}\label{tab:T=3}
	\begin{tabular}{p{.1\columnwidth}p{.15\columnwidth}p{.35\columnwidth}p{.35\columnwidth}}
		\hline\hline
		&	$|\zeta|$  & $\mathcal F~\mathrm{[\%]}$ & $\mathcal P~\mathrm{[\text{\textperthousand}]}$
		\\
		\hline
		FH &0.196 & $87.70 \pm (9.79 \times 10^{-4})$ & $0.0177 \pm (6.79 \times 10^{-5})$\\
		DH &0.173 & $99.95 \pm (6.54 \times 10^{-4})$ & $0.0011 \pm (2.81 \times 10^{-5})$\\
		\hline
		FH &0.244 & $86.58 \pm (9.89 \times 10^{-4})$ & $0.0443 \pm (1.34 \times 10^{-4})$\\
		DH &0.254 & $99.99 \pm (6.61 \times 10^{-4})$ & $0.0119 \pm (9.24 \times 10^{-5})$\\
		\hline
		FH &0.298 & $85.59 \pm (10.3 \times 10^{-4})$ & $0.1509 \pm (2.59 \times 10^{-4})$\\
		DH &0.298 & $99.98 \pm (6.63 \times 10^{-4})$ & $0.0316 \pm (1.53 \times 10^{-4})$\\
		\hline\hline
	\end{tabular}
\end{table}

\begin{table}[t!]
	\caption{%
		Fidelities and success probabilities, including their uncertainties, for $n=4$ and comparable squeezing parameters $|\zeta|$. 
	}\label{tab:T=4}
	\begin{tabular}{p{.1\columnwidth}p{.15\columnwidth}p{.35\columnwidth}p{.35\columnwidth}}
		\hline\hline
		&	$|\zeta|$  & $\mathcal F~\mathrm{[\%]}$ & $\mathcal P~\mathrm{[\text{\textperthousand}]}$
		\\
		\hline
		FH &0.167 & $74.59 \pm (14.1 \times 10^{-4})$ & $(9.14 \pm 0.58) \times 10^{-5}$\\
		DH &0.173 & $97.78 \pm (4.89 \times 10^{-4})$ & $(6.21 \pm 2.07) \times 10^{-6}$\\
		\hline
		FH &0.232 & $77.03 \pm (13.4 \times 10^{-4})$ & $(1.66 \pm 0.02) \times 10^{-3}$\\
		DH &0.254 & $99.70 \pm (5.49 \times 10^{-4})$ & $(1.11 \pm 0.08) \times 10^{-4}$\\
		\hline
		FH &0.304 & $73.32 \pm (19.8 \times 10^{-4})$ & $(7.19 \pm 0.05) \times 10^{-3}$\\
		DH &0.298 & $99.95 \pm (5.38 \times 10^{-4})$ & $(4.20 \pm 0.17) \times 10^{-4}$\\
		\hline\hline
	\end{tabular}
\end{table}

%%%%%%%%%%%%%%%%%%%%%%%%%%%%%%%%%%%%%%%%%%%%%%%%%%%%%%%%
\subsection{Nonclassicality}\label{SM:NCanalysis}

\begin{figure}[t]
	\includegraphics[width=\columnwidth]{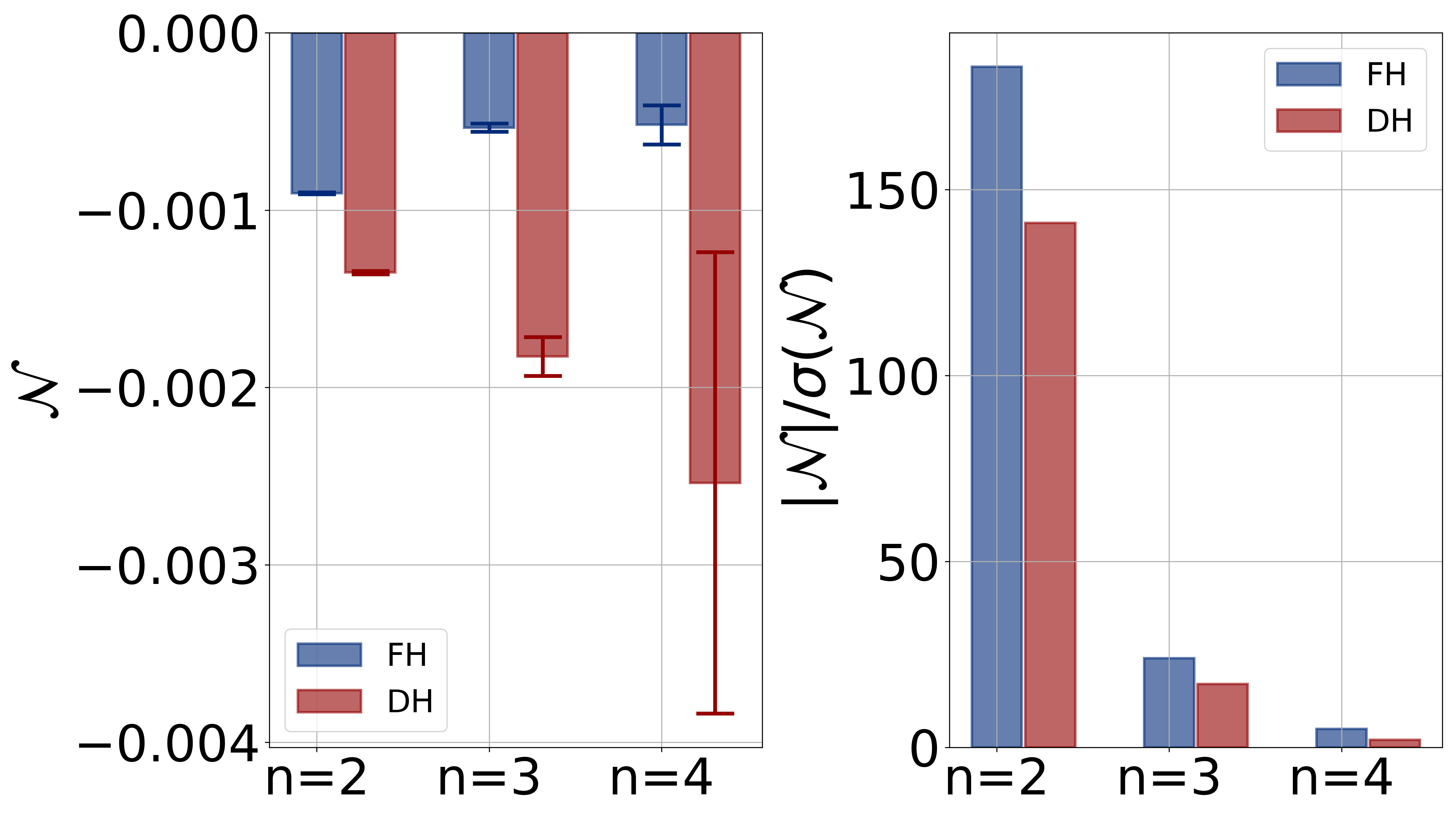}
	\includegraphics[width=\columnwidth]{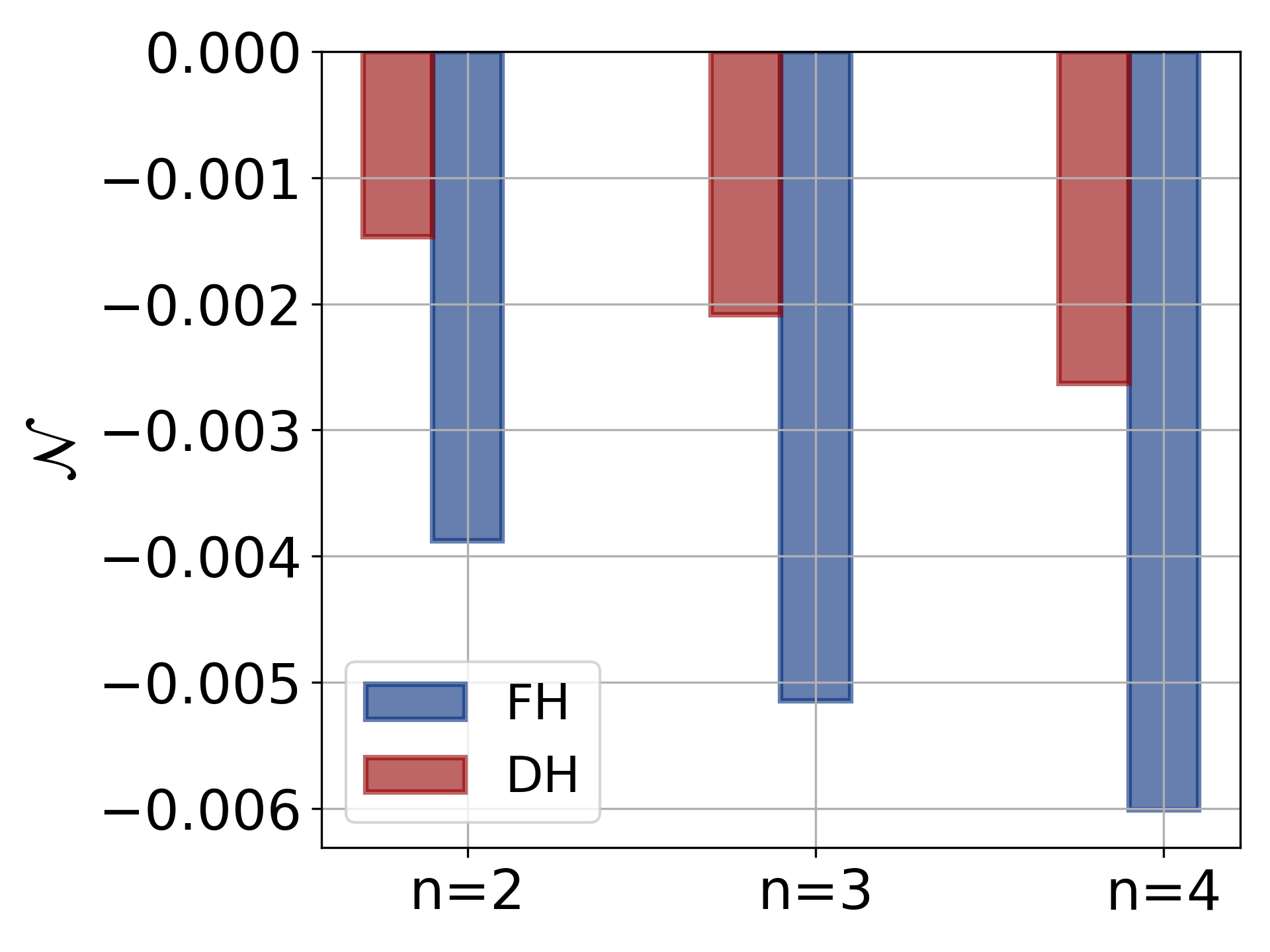}
	\caption{%
		Top:
		Nonclassicality $\mathcal{N}$ (left) and the signed significance in standard deviations of the nonclassicality $|\mathcal N|/\sigma(\mathcal{N})$ (right) for a two-, three-, and  four-photon state measured with a squeezing parameter $|\zeta|\approx 0.3$.
		Blue and red bars show the results for FH and DH, respectively.
		Bottom:
		Nonclassicality $\mathcal{N}$ simulated for the same system parameters but without loop losses.
	}\label{fig:nonclass}
\end{figure}

	Finally, we provide a closer look at the nonclassicality analysis and the effects of the loop losses.
	Figure \ref{fig:nonclass} (top) shows the presented data of the nonclassicality given in Table I in the main text for the DH and FH.
	One would naively expect that $\mathcal{N}$ is increasing with increasing photon number and that the significance is decreasing because of fewer successful events for the heralding of more photons. 
	In actually, however, the nonclassicality for the feedback-based case behaves differently since it is dropping with increasing photon numbers.
	Since the main limiting factor is the loop losses, we analysed this influence to explain this conundrum.
	Therefore, we performed a simulation of the same scenarios but without loop losses, which can be found in Fig. \ref{fig:nonclass} (bottom).
	In this scenario, we observe the expected behavior of increasing nonclassicality with increasing photon number.
	Also, the negativity $\mathcal N$ for FH is then higher than for DH.
	This further confirms that the limiting factor in the setup is the loop loss.

	Furthermore, we investigated additional click pattern for heralding multiphoton states in several round trips.
	Additional interesting cases for more than four photons, not shown in Table I in the main text, are provided in Table \ref{tab:NCSM}.

\begin{table}[b]
	\caption{%
		Nonclassicality $\mathcal{N}$ and its statistical significance $|\mathcal{N}|/\sigma(\mathcal{N})$ for different heralding click patterns, resulting in different $n$-photon states, where $n>4$, obtained from $t$ pump pulses for a squeezing parameter $|\zeta|\approx0.3$.
	}\label{tab:NCSM}
	\begin{tabular}{p{.3\columnwidth}p{.05\columnwidth}p{.05\columnwidth}p{.3\columnwidth}p{.2\columnwidth}}
		\hline\hline
		Heralding pattern & $n$ & $t$ & $\mathcal{N} \times 10^{-4}$  & $|\mathcal{N}|/\sigma(\mathcal{N})$\\
		\hline
		$(1,4)$ & 5 & 2 & $-22$\phantom{.0000}$\,\pm\,$$33$ & \phantom{00}$0.7$
		\\
		$(2,3)$ & 5 & 2 & $-17$\phantom{.0000}$\,\pm\,$$17$ & \phantom{00}$1.0$
		\\
		$(3,2)$ & 5 & 2 & $-13$\phantom{.0000}$\,\pm\,$$20$ & \phantom{00}$0.7$
		\\
		$(4,1)$ & 5 & 2 & \phantom{0}$-9$\phantom{.0000}$\,\pm\,$$32$ & \phantom{00}$0.3$
		\\
		\hline
		$(3,3)$ & 6 & 2 & $-12$\phantom{.0000}$\,\pm\,$$28$ & \phantom{00}$0.4$
		\\
		$(2,2,2)$ & 6 & 3 & \phantom{0}$-1.8$\phantom{000}$\,\pm\,$\phantom{0}$1.1$ & \phantom{00}$1.7$
		\\
		$(1,1,2,2)$ & 6 & 4 & \phantom{0}$-1.3$\phantom{000}$\,\pm\,$\phantom{0}$1.1$ & \phantom{00}$1.1$
		\\
		\hline\hline
	\end{tabular}
\end{table}

	The first four rows show different heralding click patterns that result in the heralding of $n=5$ photons in $t=2$ passes through the source.
	First, the negativity shows us that it is beneficial to have fewer photons first source passing before adding the remaining photons in the second cycle since fewer photons are subjected to round-trip losses.
	Second, the best statistical quality of $\mathcal N$ is, however, reported for the heralding click pattern $(2,3)$ since it results in more heralding detection events than the pattern $(1,4)$, requiring a four-fold coincidence.
	Thus, it is beneficial to produce the fewest photons possible per cycle and fewer photons in earlier passes for optimal nonclassicality.
	The last three rows of Table \ref{tab:NCSM} show the generation of six photons for different numbers of round trips.
	Note that the statistical significance in all given cases is expected to be quite low because of the significantly diminished number of recorded five- and six-fold coincidences over multiple round trips.

%%%%%%%%%%%%%%%%%%%%%%%%%%%%%%%%%%%%%%%%%%%%%%%%%%%%%%%%
% References
%%%%%%%%%%%%%%%%%%%%%%%%%%%%%%%%%%%%%%%%%%%%%%%%%%%%%%%%


\begin{thebibliography}{99}
	\bibitem{BB84}
		C. H. Bennett and G. Brassard,
		Quantum cryptography: Public key distribution and coin tossing,
		\href{https://doi.org/10.1016/j.tcs.2014.05.025}{Theor. Comput. Sci. \textbf{560}, 7 (2014)}.
	\bibitem{KMNRDM07}
		P. Kok, W. J. Munro, K. Nemoto, T. C. Ralph, J. P. Dowling, and G. J. Milburn,
		Linear optical quantum computing with photonic qubits,
		\href{https://doi.org/10.1103/RevModPhys.79.135}{Rev. Mod. Phys. \textbf{79}, 135 (2007)}.
	\bibitem{NC00}
		M. A. Nielsen and I. L. Chuang,
		\textit{Quantum Computation and Quantum Information}
		(\href{https://doi.org/10.1017/CBO9780511976667}{Cambridge University Press, Cambridge, England, 2000}).
	\bibitem{HB93}
		M. J. Holland and K. Burnett,
		Interferometric detection of optical phase shifts at the Heisenberg limit,
		\href{https://doi.org/10.1103/PhysRevLett.71.1355}{Phys. Rev. Lett. \textbf{71}, 1355 (1993)}
	\bibitem{OJTG07}
		A. Ourjoumtsev, H. Jeong, R. Tualle-Brouri, and P. Grangier,
		Generation of optical 'Schr\"{o}dinger cats' from photon number states,
		\href{https://doi.org/10.1038/nature06054}{Nature (London) \textbf{448}, 784 (2007)}.
	\bibitem{MPKPB19}
		E. V. Mikheev, A. S. Pugin, D. A. Kuts, S. A. Podoshvedov, and N. Ba An,
		Efficient production of large-size optical Schr\"{o}dinger cat states,
		\href{https://doi.org/10.1038/s41598-019-50703-1}{Sci. Rep. \textbf{9}, 14301 (2019)}.
	\bibitem{ZM05}
		X. Zou and W. Mathis,
		Generating a four-photon polarization-entangled cluster state,
		\href{https://doi.org/10.1103/PhysRevA.71.032308}{Phys. Rev. A \textbf{71}, 032308 (2005)}.
	\bibitem{ZPM02}
		X. Zou, K. Pahlke, and W. Mathis,
		Generation of an entangled four-photon W state,
		\href{https://doi.org/10.1103/PhysRevA.66.044302}{Phys. Rev. A \textbf{66}, 044302 (2002)}.
	\bibitem{DESBSP18}
		I. Dhand, M. Engelkemeier, L. Sansoni, S. Barkhofen, C. Silberhorn, and M. B. Plenio,
		Proposal for Quantum Simulation via All-Optically-Generated Tensor Network States,
		\href{https://doi.org/10.1103/PhysRevLett.120.130501}{Phys. Rev. Lett. \textbf{120}, 130501 (2018)}.
	\bibitem{SPBS19}
		J. Sperling, A. Perez-Leija, K. Busch, and C. Silberhorn,
		Mode-independent quantum entanglement for light,
		\href{https://doi.org/10.1103/PhysRevA.100.062129}{Phys. Rev. A \textbf{100}, 062129 (2019)}.
	\bibitem{GT07}
		N. Gisin and R. Thew,
		Quantum communication,
		\href{https://doi.org/10.1038/nphoton.2007.22}{Nat. Photonics \textbf{1}, 165 (2007)}.
	\bibitem{KLM01}
		E. Knill, R. Laflamme, and G. J. Milburn,
		A scheme for efficient quantum computation with linear optics,
		\href{https://doi.org/10.1038/35051009}{Nature (London) \textbf{409}, 45 (2001)}.
	\bibitem{VKLNFGMDS13}
		B. Vlastakis, G. Kirchmair, Z. Leghtas, S. E. Nigg, L. Frunzi, S. M. Girvin, M. Mirrahimi, M. H. Devoret, and R. J. Schoelkopf,
		Deterministically Encoding Quantum Information Using 100-Photon Schr\"{o}dinger Cat States,
		\href{https://doi.org/10.1126/science.1243289}{Science \textbf{342}, 6158 (2013)}.
	\bibitem{BFV09}
		J. O'Brien, A. Furusawa, and J. Vu\v{c}kovi\'{c},
		Photonic quantum technologies,
		\href{https://doi.org/10.1038/nphoton.2009.229}{Nat. Photonics \textbf{3}, 687 (2009)}.
	\bibitem{DMPS00}
		G. M. D'Ariano, L. Maccone, M. G. A. Paris, and M. F. Sacchi,
		Optical Fock-state synthesizer,
		\href{https://doi.org/10.1103/PhysRevA.61.053817}{Phys. Rev. A \textbf{61}, 053817 (2000)}.
	\bibitem{S05}
		K. Sanaka,
		Linear optical extraction of photon-number Fock states from coherent states,
		\href{https://doi.org/10.1103/PhysRevA.71.021801}{Phys. Rev. A \textbf{71}, 021801 (2005)}.
	\bibitem{BDSW03}
		K. R. Brown, K. M. Dani, D. M. Stamper-Kurn, and K. B. Whaley,
		Deterministic optical Fock-state generation,
		\href{https://doi.org/10.1103/PhysRevA.67.043818}{Phys. Rev. A \textbf{67}, 043818 (2003)}.
	\bibitem{MK09}
		K. T. McCusker and P. G. Kwiat,
		Efficient Optical Quantum State Engineering,
		\href{https://doi.org/10.1103/PhysRevLett.103.163602}{Phys. Rev. Lett. \textbf{103}, 163602 (2009)}.
	\bibitem{GFM14}
		B. L. Glebov, J. Fan, and A. Migdall,
		Photon number squeezing in repeated parametric downconversion with ancillary photon-number measurements,
		\href{https://doi.org/10.1364/OE.22.020358}{Opt. Express \textbf{22}, 20358 (2014)}.
	\bibitem{BBFKT19}
		M. Bouillard, G. Boucher, J. Ferrer Ortas, B. Kanseri, and R. Tualle-Brouri,
		High production rate of single-photon and two-photon Fock states for quantum state engineering,
		\href{https://doi.org/10.1364/OE.27.003113}{Opt. Express \textbf{27}, 3113 (2019)}.
	\bibitem{ZVB04}
		A. Zavatta, S. Viciani, and M. Bellini,
		Tomographic reconstruction of the single-photon Fock state by high-frequency homodyne detection,
		\href{https://doi.org/10.1103/PhysRevA.70.053821}{Phys. Rev. A \textbf{70}, 053821 (2004)}.
	\bibitem{YMKLF13}
		J. Yoshikawa, K. Makino, S. Kurata, P. van Loock, and A. Furusawa,
		Creation, Storage, and On-Demand Release of Optical Quantum States with a Negative Wigner Function,
		\href{https://doi.org/10.1103/PhysRevX.3.041028}{Phys. Rev. X \textbf{3}, 041028 (2013)}.
	\bibitem{NALDT15}
		L. A. Ngah, O. Alibart, L. Labont\'{e}, V. D'Auria, and S. Tanzilli,
		Ultra‐fast heralded single photon source based on telecom technology,
		\href{https://doi.org/10.1002/lpor.201400404}{Laser Photonics Rev. \textbf{9}, L1 (2015)}.
	\bibitem{CWSS13}
		M. Cooper, L. J. Wright, C. S\"{o}ller, and B. J. Smith,
		Experimental generation of multiphoton Fock states,
		\href{https://doi.org/10.1364/OE.21.005309}{Opt. Express \textbf{21}, 5309 (2013)}.
	\bibitem{WDY06}
		E. Waks, E. Diamanti, and Y. Yamamoto,
		Generation of photon number states,
		\href{https://doi.org/10.1088/1367-2630/8/1/004}{New J. Phys. \textbf{8}, 4 (2006)}.
	\bibitem{TBHLNGS19}
		J. Tiedau, T. J. Bartley, G. Harder, A. E. Lita, S. Woo Nam, T. Gerrits, and C. Silberhorn,
		Scalability of parametric down-conversion for generating higher-order Fock states,
		\href{https://doi.org/10.1103/PhysRevA.100.041802}{Phys. Rev. A \textbf{100}, 041802 (2019)}.
	\bibitem{MSM20}
		E. Meyer-Scott, C. Silberhorn, and A. Migdall,
		Single-photon sources: Approaching the ideal through multiplexing featured,
		\href{https://doi.org/10.1063/5.0003320}{Rev. Sci. Instrum. \textbf{91}, 041101 (2020)}.
	\bibitem{KK19}
		F. Kaneda and P. G. Kwiat,
		High-efficiency single-photon generation via large-scale active time multiplexing,
		\href{https://doi.org/10.1126/sciadv.aaw8586}{Sci. Adv. \textbf{5}, eaaw8586 (2019)}.
	\bibitem{BDSJBDSW12}
		T. J. Bartley, G. Donati, J. B. Spring, X.-M. Jin, M. Barbieri, A. Datta, B. J. Smith, and I. A. Walmsley,
		Multiphoton state engineering by heralded interference between single photons and coherent states,
		\href{https://doi.org/10.1103/PhysRevA.86.043820}{Phys. Rev. A \textbf{86}, 043820 (2012)}.
	\bibitem{ELSDBDPS20}
		M. Engelkemeier, L. Lorz, S. De, B. Brecht, I. Dhand, M. B. Plenio, C. Silberhorn, and J. Sperling,
		Quantum photonics with active feedback loops,
		\href{https://doi.org/10.1103/PhysRevA.102.023712}{Phys. Rev. A. \textbf{102}, 023712 (2020)}.
%%%%%%%%%%%%%%%%%%%%%%%%%%%%%%%%%%%%%%%%%%%%%%%%%%%%
	\bibitem{PTKJ96}
		H. Paul, P. T\"orm\"a, T. Kiss, and I. Jex,
		Photon Chopping: New Way to Measure the Quantum State of Light,
		\href{https://doi.org/10.1103/PhysRevLett.76.2464}{Phys. Rev. Lett. \textbf{76}, 2464 (1996)}.
	\bibitem{ASSBWFJPF04}
		D. Achilles, C. Silberhorn, C. Sliwa, K. Banaszek, I. A. Walmsley, M. J. Fitch, B. C. Jacobs, T. B. Pittman, and J. D.Franson,
		Photon-number-resolving detection using time-multiplexing,
		\href{https://doi.org/10.1080/09500340408235288}{J. Mod. Opt. \textbf{51}, 1499 (2004)}.
	\bibitem{SVA12}
		J. Sperling, W. Vogel, and G. S. Agarwal,
		True photocounting statistics of multiple on-off detectors,
		\href{https://doi.org/10.1103/PhysRevA.85.023820}{Phys. Rev. A \textbf{85}, 023820 (2012)}.
	\bibitem{ECMS11}
		A. Eckstein, A. Christ, P. J. Mosley, and C. Silberhorn,
		Highly Efficient Single-Pass Source of Pulsed Single-Mode Twin Beams of Light,
		\href{https://doi.org/10.1103/PhysRevLett.106.013603}{Phys. Rev. Lett. \textbf{106}, 013603 (2011)}.
	\bibitem{comment:Squeezing}
		The two-mode squeezing operator reads as $\hat S=\exp(\zeta\hat a^\dag\hat b^\dag-\zeta^\ast\hat a\hat b)$, where $\zeta$ is referred to as the squeezing parameter and is proportional to the product of the interaction time, coupling strength, and the field amplitude of the pump, i.e., the square root of the pump power, $|\zeta|\propto \sqrt{P}$.
		The squeezing in decibel is given by $-10\log_{10}(e^{-2|\zeta|})=20\log_{10}(e)\times|\zeta|$.
	\bibitem{SVA13}
		J. Sperling, W. Vogel, and G. S. Agarwal,
		Correlation measurements with on-off detectors,
		\href{https://doi.org/10.1103/PhysRevA.88.043821}{Phys. Rev. A \textbf{88}, 043821 (2013)}.
	\bibitem{SBVHBAS15}
		J. Sperling, M. Bohmann, W. Vogel, G. Harder, B. Brecht, V. Ansari, and C. Silberhorn,
		Uncovering Quantum Correlations with Time-Multiplexed Click Detection,
		\href{https://doi.org/10.1103/PhysRevLett.115.023601,}{Phys. Rev. Lett. \textbf{115}, 023601 (2015)}.
	\bibitem{comment:SupplementalMaterial}
		See Supplemental Material %, which includes Refs. ...,
		for additional details about theory and modeling, data processing and error propagation, and additional results.

\end{thebibliography}
\end{document}